\documentclass[11pt,a4paper]{article}
\pdfoutput=1
\usepackage{jheppub}
\usepackage{mathtools}
\usepackage[all]{hypcap}
\usepackage{xcolor}
\usepackage{url}
\usepackage[utf8]{inputenc}
\allowdisplaybreaks%

\newcommand{\be}{\begin{equation}}
\newcommand{\ee}{\end{equation}}
\newcommand{\bea}{\begin{eqnarray}}
\newcommand{\eea}{\end{eqnarray}}
\newcommand{\beq}{\begin{equation}}
\newcommand{\eeq}{\end{equation}}
\newcommand{\comment}[1]{}

\begin{document}
 
{\raggedleft PNUTP-19-A13\\}

\title{Freeze-in Axion-like Dark Matter}

\affiliation{Department of Physics, Pusan National University, Busan 46241, Korea}
 
\author{Sang~Hui~Im,}
\author{Kwang~Sik~Jeong} 
 
\emailAdd{imsanghui@pusan.ac.kr}
\emailAdd{ksjeong@pusan.ac.kr} 

\abstract{
We present an interesting Higgs portal model where an axion-like particle (ALP) couples to 
the Standard Model sector only via the Higgs field. 
The ALP becomes stable due to CP invariance and turns out to be a natural candidate 
for freeze-in dark matter because its properties are controlled by the perturbative 
ALP shift symmetry. 
The portal coupling can be generated non-perturbatively by a hidden confining gauge 
sector, or radiatively by new leptons charged under the ALP shift symmetry.
Such UV completions generally involve a CP violating phase, which makes the ALP unstable 
and decay through mixing with the Higgs boson, but can be sufficiently suppressed 
in a natural way by invoking additional symmetries.  
}

\maketitle
 
\section{Introduction}

The existence of dark matter in the universe is a convincing evidence for physics beyond 
the Standard Model (SM) of particle physics.
A weakly interacting massive particle (WIMP) has been considered as an attractive candidate for 
cold dark matter because it appears often and naturally if one extends the SM to address 
its puzzles such as the smallness of the electroweak 
scale~\cite{Arcadi:2017kky, Roszkowski:2017nbc}. 
However, the negative results so far both in direct and indirect detections of a WIMP dark matter
lead to consider alternative dark matter scenarios. 

On the other hand, the LHC experiments have given no clear signals for new physics beyond the SM,
and showed only small deviations of Higgs couplings from their SM values. 
This may indicate that a dark matter particle couples very weakly to the SM particles, for which
its relic density generally depends on the history of the early universe.
An axion-like particle (ALP) is an appealing candidate for such feebly interacting dark matter since
the associated shift symmetry controls its mass and couplings.

If added to extend the SM, an ALP is usually considered to have an anomalous coupling to SM gauge
bosons, and is subject to various experimental constraints depending on its mass and decay
constant.
For instance, anomalously coupled to gluons, the QCD axion solves the strong CP 
problem~\cite{PhysRevLett.38.1440, Kim:2008hd},
and it can also make up a significant fraction of dark matter as produced by 
the misalignment mechanism or from topological defects.
Another possibility, which has recently attracted growing interest, is that an ALP couples 
to the SM sector via a Higgs portal.
Such an ALP can play a crucial role in electroweak phase transition, possibly providing an explanation
why the electroweak scale is small~\cite{Graham:2015cka, Choi:2015fiu, Kaplan:2015fuy}, 
or how the baryon asymmetry of the universe is produced~\cite{Son:2018avk, Jeong:2018ucz, Jeong:2018jqe, Abel:2018fqg, Gupta:2019ueh}.

In this paper we explore if the dark matter of the universe can be explained by an ALP which
couples to the SM sector only via a Higgs portal.
The ALP is stabilized at a CP conserving minimum and becomes stable if it has no other non-derivative 
interactions. 
The portal coupling should then be tiny to make the ALP not thermalized with SM particles
since otherwise it would overclose the universe in most of the parameter space satisfying 
the experimental constraints on dark matter scattering with nuclei.
In such a case, ALP dark matter can still be generated out of equilibrium via 
the freeze-in mechanism~\cite{McDonald:2001vt, Choi:2005vq, Hall:2009bx, Bernal:2017kxu}.
It should be noted that the ALP shift symmetry can naturally suppress the portal coupling 
so that Higgs properties are rarely modified and the freeze-in occurs to produce the correct dark matter
abundance.

The portal coupling can be generated non-perturbatively from a hidden confining gauge sector,
or radiatively by new leptons which have shift-symmetric interactions with the Higgs field.  
In UV completions of the Higgs portal, one generally encounters a CP violating phase that 
makes the ALP unstable and decay to SM particles through mixing with the Higgs boson. 
Such harmful coupling can be sufficiently suppressed in a natural way by invoking supersymmetry.
The ALP is never in thermal equilibrium as long as extra SM-charged particles responsible for 
a portal coupling are heavier than the reheating temperature of inflation,
or if ALP interactions are weak enough.

This paper is organized as follows. 
In section \ref{ALP-DM} we introduce a novel model for freeze-in dark matter where an ALP interacts 
with the SM only via the Higgs field, and examine the parameter space where the ALP constitute 
all the dark matter of the universe. 
Section \ref{uv-completion} is devoted to discussions on a UV completion of the portal coupling
and its cosmological aspects. 
The final section is for conclusions.

\section{Higgs Portal with Axion-like Dark Matter}
\label{ALP-DM}

Having properties controlled by the associated shift symmetry, an ALP coupled to the SM via 
a Higgs portal becomes a natural candidate for feebly interacting dark matter.
In this section we construct a portal model where the production of ALP dark matter takes
place through freeze-in.

\subsection{Higgs Portal}
 
Coupling to the SM sector via a Higgs portal, an ALP can cosmologically relax the Higgs boson mass to 
the weak scale~\cite{Graham:2015cka, Choi:2015fiu, Kaplan:2015fuy}, or can drive a strong first-order phase transition required to implement
electroweak baryogenesis~\cite{Jeong:2018ucz, Jeong:2018jqe}.  
In this paper we explore the possibility that the dark matter of the universe is explained by
an ALP with a Higgs portal 
\bea
\label{Higgs-ALP-coupling}
-M^2 \cos\left(\frac{\phi}{f}\right) |H|^2,
\eea
but without any anomalous coupling to gauge bosons or derivative couplings below a cutoff scale $\Lambda$.
Here $H$ is the Higgs doublet field, and $f$ is the energy scale at which there occurs
a transition from linear to non-linear phase associated with the ALP shift symmetry.  
UV completions of the model will be presented in section~\ref{uv-completion}.

The ALP has a non-derivative coupling to the gauge-invariant Higgs squared operator,
and thus its potential is radiatively generated from Higgs loops but without introducing a new 
CP violating phase.  
The scalar potential is thus written
\bea
V = \lambda |H|^4 + \mu^2 |H|^2 - M^2 \cos\left(\frac{\phi}{f} \right) |H|^2
- \frac{1}{16\pi^2} M^2 \Lambda^2 \cos\left(\frac{\phi}{f} \right),
\eea
assuming the absence of other effects breaking the ALP shift symmetry explicitly. 
Here the last term arises from a closed Higgs loop, and for simplicity we have absorbed a coefficient 
of order unity into the effective cutoff scale $\Lambda$.  
Note that $f$ should be larger than $\Lambda$ for the consistency of the effective theory.

It is clear that the ALP is stabilized at a CP preserving minimum $\phi=0$ 
and becomes stable due to $Z_2$ symmetry, $\phi \to -\phi$.
The ALP can therefore contribute to the dark matter of the universe. 
Around the minimum $\phi=0$, the potential is expanded as
\bea
V = \lambda |H|^4 + (\mu^2 - M^2) |H|^2
+ \frac{1}{2} \lambda_{h\phi} |H|^2 \phi^2  
+ \frac{1}{2} m^2_\phi \phi^2 + \cdots, 
\eea
where the ellipsis includes higher dimensional operators, and
the ALP mass and couplings are given by  
\bea
\lambda_{h\phi} &=& \left(\frac{M}{f}\right)^2,
\nonumber \\ 
m_\phi  &=& 
\sqrt{ 1+ (4\pi v/\Lambda)^2} \times
\frac{ \sqrt{\lambda_{h\phi} }}{4\pi} \Lambda,  
\eea
with $v\simeq174$~GeV being the Higgs vacuum expectation value. 
Note that Higgs mixing with the ALP is subject to various experimental 
constraints~\cite{Choi:2016luu, Flacke:2016szy}.
In our model, however, the neutral Higgs boson $h$ does not mix with $\phi$, implying that
the Higgs couplings are not modified.  
The LHC data on the $125$~GeV Higgs boson constrain only the coupling $\lambda_{h\phi}$ if 
a Higgs invisible decay to ALPs is open.

If produced by thermal freeze-out, ALP dark matter overcloses the universe 
in most of the parameter space allowed by  
the experimental limit on spin-independent interactions with nuclei and
LHC searches for Higgs invisible decays~\cite{Hardy:2018bph}. 
To prevent the ALP from thermalizing with the SM plasma, one needs to make it feebly
interact with the SM sector with~\cite{Enqvist:2014zqa} 
\bea
\lambda_{h\phi} \lesssim 10^{-7}.
\eea 
In our scenario, this can be achieved in a natural way because all the ALP couplings are controlled by 
the scale $f$. 
The ALP never thermalizes, for instance, for $M$ around the weak scale if $f$ is above $10^6$~GeV.

\subsection{Freeze-in Production}

Even in the case that dark matter was never in thermal equilibrium with the SM plasma, it can be 
produced from decays or scatterings of thermal particles and make up a sizable fraction
of the observed dark matter density~\cite{Hall:2009bx, Bernal:2017kxu}.
We note that the Higgs portal with an ALP provides a natural framework for such freeze-in dark matter.
The ALP comoving abundance freezes to a constant value as the Higgs boson number density 
is Boltzmann-suppressed to stop producing ALPs or the temperature of 
the universe drops below the ALP mass.

The time evolution of the ALP number density is described by a Boltzmann equation.
In the Higgs portal scenario, the ALP couples to the SM sector only via the interactions  
\bea
\frac{\lambda_{h\phi}}{4}  h^2  \phi^2
+ \frac{\lambda_{h\phi} v}{\sqrt2}  h \phi^2
\eea
with a tiny portal coupling $\lambda_{h\phi}=(M/f)^2\lesssim 10^{-7}$ to avoid overclosure 
of the universe.
Then ALP dark matter does not reach thermal equilibrium, but is produced via the processes 
$h \to \phi\phi$ and $hh\to \phi\phi$ depending on $m_\phi$, $m_h$ and $\lambda_{h\phi}$.
The dominant freeze-in process is via decays of Higgs bosons in thermal bath if kinematically allowed.
Under the assumption that the initial abundance is negligible and the number of relativistic degrees of
freedom does not change during ALP production, 
the approximate solution for the ALP relic abundance is found to be~\cite{Hall:2009bx}
\bea
\Omega_\phi h^2 &\simeq& 
\left.  \frac{2.2\times 10^{27} }{g_{\ast s}\sqrt{g_\ast} }
\frac{m_\phi\,\Gamma(h\to \phi\phi)}{m^2_h}\right|_{T\simeq m_h}
\nonumber \\
&\simeq& 
8.8 \times 10^{25}
\sqrt{1-\frac{4m^2_\phi}{m^2_h} }
\frac{m_\phi v^2}{m^3_h}
\left.\frac{\lambda^2_{h\phi}}{ g_{\ast s} \sqrt{g_\ast}}\,\right|_{T\simeq m_h},
\eea
for $2 m_\phi<m_h$ with $\Gamma(h\to ii)$ being the Higgs decay width of the indicated mode,
while it is given by 
\bea
\Omega_\phi h^2 
&\simeq&
2.2\times 10^{23} 
\left.\frac{\lambda^2_{h\phi}}{ g_{\ast s} \sqrt{g_\ast}}\,\right|_{T\simeq m_\phi},
\eea
for $2 m_\phi>m_h$.
Here $g_{\ast s}$ and $g_{\ast}$ are the effective numbers of degrees of freedom related to 
the entropy and energy densities, respectively.  
 
Using the approximate solution, one can examine in which parameter space the freeze-in ALP 
accounts for the observed dark matter density. 
For the cutoff scale $\Lambda$ above TeV, the ALP mass is given by 
$m_\phi \simeq \sqrt{\lambda_{h\phi}}\Lambda/4\pi$ with $\lambda_{h\phi}=(M/f)^2$,
and it is found that the correct dark matter density is obtained when the ALP has
\bea
\lambda_{h\phi} 
&\simeq&  10^{-10} 
\left(\frac{g_{\ast s}\sqrt{g_\ast}}{10^3}\right)^{2/5}
\left(\frac{\Lambda}{10^3 {\rm GeV}}\right)^{-2/5},
\nonumber \\
m_\phi 
&\simeq& 
1\,{\rm MeV}
\left(\frac{g_{\ast s}\sqrt{g_\ast}}{10^3}\right)^{1/5}
\left(\frac{\Lambda}{10^3{\rm GeV}}\right)^{4/5}, \label{ALPmass1}
\eea 
if $2m_\phi<m_h$, and 
\bea
\lambda_{h\phi}
&\simeq&
2\times 10^{-11}
\left(\frac{g_{\ast s}\sqrt{g_\ast}}{10^3}\right)^{1/2},
\nonumber \\
m_\phi 
&\simeq& 
380\,{\rm GeV} 
\left(\frac{g_{\ast s}\sqrt{g_\ast}}{10^3}\right)^{1/4}
\left( \frac{\Lambda}{10^9{\rm GeV}}\right), \label{ALPmass2}
\eea
otherwise.
The above shows that there are two viable regions of parameter space
depending on whether $h\to \phi\phi$ is allowed or not:
\bea \label{par_space}
&\mbox{(a)}& 
f  \sim 
10^8\,{\rm GeV} 
\left(\frac{M}{10^3{\rm GeV}}\right) 
\left( \frac{\Lambda}{10^3{\rm GeV}} \right)^{1/5}
\quad{\rm and}\quad
\Lambda \lesssim 2\times 10^8{\rm GeV},
\nonumber \\
&\mbox{(b)}&  
f \sim
2\times 10^8\,{\rm GeV}
\left(\frac{M}{10^3{\rm GeV}}\right)
\quad{\rm and}\quad
\Lambda \gtrsim 2\times 10^8{\rm GeV}, 
\eea  
where we have taken $g_{\ast s} \sqrt{g_\ast}=10^3$.
In the region (b), where Higgs decay to ALPs is kinematically forbidden, 
the consistency condition $f>\Lambda$ indicates that $M$ is above TeV.  
In order to  avoid large fine-tuning in electroweak symmetry breaking, one may 
constrain $M$ to be not hierarchically larger than the weak scale. 
We also note that $\Omega_\phi h^2 =  0.12$ requires $m_\phi \gtrsim 100$~keV for $\Lambda$ above
the weak scale, for which the ALP dark matter is sufficiently cold to form structures of the universe.  
If light, the ALP can contribute to dark radiation during big-bang nucleosynthesis (BBN) but only with
$\Delta N_{\rm eff}$ below order $10^{-5}$ for $m_\phi$ above $100$~keV.

We close this subsection by discussing how much ALP coherent oscillations can contribute to
the dark matter abundance.
The ALP field starts to oscillate about the minimum at $T=T_{\rm osc}$ when the expansion 
rate of the universe becomes comparable to its mass. 
Using the fact that ALP field oscillation behaves like cold dark matter, one can estimate
the relic abundance to be
\bea
\Omega_\phi h^2 \simeq 
0.5\times 10^{-4}  
\left( \frac{g_\ast(T_{\rm osc})}{100} \right)^{-1/4}
\left(\frac{m_\phi}{1{\rm MeV}} \right)
\left(\frac{f}{10^8 {\rm GeV}}\right)^2 
\left(\frac{T_{\rm osc}}{10^7{\rm GeV}} \right)^{-1} 
\theta^2_{\rm ini},
\eea
where $\theta_{\rm ini}=\phi_{\rm ini}/f$ is the initial misalignment angle,
and $T_{\rm osc}$ cannot exceed $f$ because the ALP appears at energy 
scales below $f$.
The contribution from coherent oscillations is thus negligible if 
\bea
T_{\rm osc} \gg
5\times 10^2 {\rm GeV}
\left( \frac{m_\phi}{1{\rm MeV}} \right)
\left( \frac{f}{10^8{\rm GeV}} \right)^2, 
\eea
taking $g_\ast=100$.
Combined with $T_{\rm osc}<f$, the above relation shows that coherent oscillations can 
constitute a sizable fraction of the observed dark matter density if
$m_\phi$ is around or above $200{\rm GeV}\times(10^8{\rm GeV}/f)$, 
i.e.~in some part of the parameter space for freeze-in if $\theta_{\rm ini}$
is of order unity.
Note that if the ALP mass does not depend on temperature, $T_{\rm osc}$ is given
by $T_{\rm osc}\simeq 2\times 10^7 {\rm GeV} (m_\phi/{\rm MeV})^{1/2}$,
and thus is high enough to suppress the contribution from coherent oscillations in 
most of the parameter space for freeze-in. 
On the other hand, if non-perturbatively generated by some hidden confining gauge 
interaction,
the ALP mass is turned on and grows only at temperatures around the confinement scale,
implying that $T_{\rm osc}$ is determined by the confinement scale, depending on
how much the hidden sector is colder than the SM plasma.

\subsection{Longevity}

If high energy dynamics generating the Higgs portal coupling involves a CP violating phase, 
there generally appears an effective potential of the form 
\bea
\Delta V = -\mu^4_\phi \cos\left( \frac{\phi}{f} + \alpha \right),
\eea
which shifts the minimum to a CP violating point, and consequently  
the ALP mixes with the Higgs boson and couples to other SM particles.  
Let us consider the case with $\mu^4_\phi \ll M^2\Lambda^2/16\pi^2$, for which 
$\Delta V$ gives only a negligible contribution to the ALP mass.
In such case, the mixing angle between $\phi$ and $h$ is estimated by
\bea
\theta_{\rm mix}  
\simeq
\frac{16\pi^2 \mu^4_\phi}{(m^2_h - m^2_\phi)\Lambda^2}\frac{v}{f}
\sin\alpha,
\eea
and it should be small enough 
in order for the ALP to live longer than the age of the universe, i.e.~to ensure
$\tau > 5\times 10^{17}\,{\rm s}$
where $\tau$ is the ALP lifetime. 
For $m_\phi$ in the range for freeze-in, the mixing is more severely constrained by 
gamma ray observations because it makes the ALP directly decay into 
photons and possibly into other SM particles that  subsequently produce photons via inverse Compton 
scattering~\cite{Cadamuro:2011fd, Essig:2013goa, Perez:2016tcq,  Ackermann:2015lka, Blanco:2018esa}.
The current experimental bound on $\tau$ ranges from about $10^{27}\,{\rm s}$ to $10^{29}\,{\rm s}$
depending on the ALP mass. 
Here we shall conservatively take a lower bound on the ALP lifetime to be 
$\tau>10^{28}\,{\rm s}$ for the ALP in the MeV to GeV range, and 
$\tau>10^{29}\,{\rm s}$ if above GeV. 

For $m_\phi <2m_\mu$ with $m_\mu$ being the muon mass, 
the ALP dominantly decays into electrons via mixing with the Higgs boson,
and its lifetime should be long enough to evade the constraints from galactic photon spectra.
The longevity bound on $\theta_{\rm mix}$ requires
\bea
\label{longevity-1}
\frac{\mu^4_\phi\sin\alpha}{M\Lambda^3} &\lesssim&
10^{-18}
\left(\frac{m_\phi}{10\,{\rm MeV}}\right)^{-3/2} \left(\frac{\tau}{10^{28}\,{\rm s}}\right)^{-1/2},
\eea
where we have used the relation $f \simeq M\Lambda/(4\pi m_\phi)$, which 
holds for $\Lambda$ above TeV.
If the ALP is heavy and decays into other SM particles,
the constraint on the mixing gets severer.  
For instance, the longevity bound reads
\bea
\label{longevity-2}
\frac{\mu^4_\phi\sin\alpha}{M\Lambda^3} &\lesssim& 
 10^{-30}
\left(\frac{m_\phi}{10^3\,{\rm GeV}}\right)^{-1/2} \left(\frac{\tau}{10^{29}\,{\rm s}}\right)^{-1/2},
\eea 
if $m_\phi \gtrsim 140$~GeV, for which the ALP dominantly decays to $W$ bosons.

\section{UV Completion}
\label{uv-completion}

In this section we discuss how to UV complete the Higgs portal model. 
The UV completion should be such that CP violation causing mixing between
the ALP and the Higgs boson if any is sufficiently suppressed,
and the ALP is not thermalized with SM particles. 
This can be achieved with help of perturbative ALP shift symmetry and supersymmetry. 

\subsection{Non-perturbative Higgs Portal}

The portal coupling (\ref{Higgs-ALP-coupling}) can be induced nonperturbatively 
if ALP shift symmetry is anomalously broken by hidden QCD confining at $\Lambda_{\rm hid}$.
Let us introduce vector-like lepton doublets $L+L^c$ and singlets $N+N^c$ which are charged 
under the hidden QCD and have interactions preserving the ALP shift symmetry.
One can perform appropriate field redefinitions to write their interactions
without loss of generality as
\bea
\label{effective-model}
y H L N^c + y^\prime H^\dagger L^c N + m_L LL^c + \mu_N e^{i\alpha} NN^c,
\eea 
in which the Yukawa couplings and mass parameters are all real and positive.  
For the case with $\mu_N<\Lambda_{\rm hid} < m_L$, the singlet leptons have effective 
interactions
\bea
\frac{yy^\prime}{m_L} |H|^2 NN^c
+ \left( \mu_N e^{i\alpha} + \frac{yy^\prime}{16\pi^2} m_L 
\ln\left(\frac{\Lambda^2_\ast}{m^2_h}\right) \right)NN^c,  
\eea
after integrating out heavy lepton doublets.
Here $\Lambda_\ast$ is the cutoff scale of the UV model, and 
we have included radiative contributions to the lepton singlet mass
arsing from loops of the Higgs and lepton doublets.  
If the ALP has an anomalous coupling to the hidden QCD, 
the scalar potential is written
\bea
V = \lambda |H|^4 + \mu^2 |H|^2 + \Delta V_{\rm eff},
\eea
at energy scales below $\Lambda_{\rm hid}$,
where $\Delta V_{\rm eff}$ is the effective potential obtained by integrating out 
the heavy meson field from $N+N^c$ condensation
\bea
\label{NP-Veff}
\Delta V_{\rm eff} = 
- M^2\cos\left(\frac{\phi}{f}\right) |H|^2
- \frac{1}{16\pi^2}M^2\Lambda^2 \cos\left(\frac{\phi}{f}\right)
- \mu^4_\phi \cos\left(\frac{\phi}{f} + \alpha \right),
\eea
in which the involved parameters are determined by the couplings of high energy theory as
follows
\bea
\label{hidden-QCD-model}
M^2 &=& \frac{yy^\prime \Lambda^3_{\rm hid}}{m_L},
\nonumber \\
\Lambda^2 &=& m^2_L 
\ln\left(\frac{\Lambda^2_\ast}{m^2_h }\right),
\nonumber \\
\mu^4_\phi &=&
\mu_N \Lambda^3_{\rm hid}.
\eea 
The effective potential assumes that the lepton doublets are heavier than the confinement
scale while the lepton singlets are lighter, implying that $\Lambda_{\rm hid}$
lies in the range
\bea
\label{Confinement-condition}
\frac{yy^\prime}{16\pi^2}m_L \ln\left(\frac{\Lambda^2_\ast}{m^2_h}\right)
< \Lambda_{\rm hid} < m_L,
\eea
for the lepton singlets with a mass dominated by the radiative contribution. 
We note that the portal coupling is controllable in the sense that it vanishes
in the limit that the gauge coupling of hidden QCD goes to zero.

The CP violating term in the effective potential should be highly suppressed because it
makes the ALP unstable and decay into SM particles via mixing with the Higgs boson. 
One way would be to impose that the hidden QCD sector preserves CP invariance, 
for which $\alpha=0$.  
Another way is to invoke supersymmetry to suppress the mass parameter $\mu_N$.  
This can be achieved if the leptons couple to a singlet scalar $X$ to 
acquire their masses after spontaneous U$(1)_X$ breakdown in such a way that  
$m_L$ arises from superpotential while $\mu_N$ is from K\"ahler potential.
Here U$(1)_X$ symmetry, which can be either global or gauged,  
is assumed to be spontaneously broken at a scale lower than $f$.
If global, we further assume that some hidden confining dynamics makes 
the phase component of $X$ much heavier than the ALP. 
To promote U$(1)_X$ to a gauge symmetry,
one should include additional fermions to satisfy anomaly cancellation.

As an explicit example with global U$(1)_X$,
let us consider a simple model where the operators responsible for lepton masses are given by 
\bea
\label{susy-model}
K &\ni&  \frac{X^*}{M_{Pl}}NN^c + {\rm h.c.},
\nonumber \\
W &\ni& XLL^c + H_d LN^c + H_u L^c N,
\eea 
omitting dimensionless coupling constants,
in the framework of flavor and CP conserving mediation of supersymmetry breaking,\footnote{
A Dirac mass for the singlet leptons $N+N^c$ is radiatively generated from the loops
formed by the Higgs scalars and doublet leptons, and also from the loops of their superpartners. 
Those would make the ALP unstable if the soft supersymmetry breaking parameters associated 
with $H_uH_d$ and $XLL^c$ introduce a new CP violating phase, i.e.~a CP phase that cannot be 
rotated away by field redefinitions. 
}
such as gauge, anomaly, and moduli mediation, so that soft supersymmetry breaking terms 
do not contain new sources of flavor and CP violation.\footnote{
Note that there are gravity mediation effects from non-renormalizable Planck-suppressed operators, 
which generally induce CP violations and thus should be highly suppressed to satisfy 
the longevity conditions.
Using the relations (\ref{hidden-QCD-model}) and (\ref{Confinement-condition}), 
one finds that the condition (\ref{longevity-1}) 
requires 
\bea
\frac{\epsilon m_{3/2}}{m_{\rm SUSY}}
\lesssim
\frac{10^{-12}}{(yy^\prime)^2} 
\left( \frac{m_\phi}{10{\rm MeV}}\right)^{-3/2} 
\left(\frac{\tau}{10^{28}{\rm s}}\right)^{-1/2},
\nonumber
\eea
for $m_\phi < 2m_\mu$, barring an alignment of the associated CP violating phases.
Here $m_{3/2}$ is the gravitino mass, and 
$\epsilon$ represents the degree of sequestering between the visible sector and the supersymmetry 
breaking sector.
The above indicates that the ALP becomes a proper dark matter candidate
if $\epsilon\ll 1$ and/or $yy^\prime \ll 1$, i.e.~if the sequestering is 
strong and/or the Yukawa couplings are small.  
In gauge mediation, the constraint is weaker than in other mediation schemes because
 the gravitino 
can be very light,
\bea
\frac{(16\pi^2)^2 m_{\rm SUSY}}{M_{Pl}}
< \frac{m_{3/2}}{m_{\rm SUSY}} \ll 1, 
\nonumber
\eea 
for the messenger scale higher than $16\pi^2 m_{\rm SUSY}$ as required to give non-tachyonic masses
to the messenger scalars.  
}
Here $M_{Pl}$ denotes the reduced Planck mass, and 
$H_u$ and $H_d$ are the conventional Higgs doublet superfields which are singlets
under U$(1)_X$ and ALP shift transformations. 
The involved superfields carry charges given by
\bea 
\begin{tabular}{| c | c | c | c | c | c | c | c |} 
\hline
            & $H_u$ & $H_d$   & $L$   &  $L^c$    & $N$  & $N^c$   & $X$         \\ \hline       
SU$(2)_L$  & ${\bf 2}$ & ${\bf \bar 2}$   & ${\bf 2}$      &  ${\bf \bar 2}$    & ${\bf 1}$  & ${\bf 1}$     & ${\bf 1}$ \\ \hline  
U$(1)_Y$   & $+1/2$ & $-1/2$  & $+1/2$              & $-1/2$             & $0$     & $0$          & $0$ \\ \hline  
SU$(N)$  & ${\bf 1}$ & ${\bf 1}$ & ${\bf N}$  & ${\bf \bar N}$   
 & ${\bf N}$             &  ${\bf \bar N}$  & ${\bf 1}$ \\ \hline  
U$(1)_X$  &0 & 0  & $-1/2$              & $-1/2$             & $+1/2$     & $+1/2$          & $+1$ \\ \hline  
\end{tabular}  
\nonumber 
\eea
taking SU$(N)$ for a hidden confining gauge group.  
From the above interactions, which preserve both U$(1)_X$ and ALP shift symmetries, it
is found that spontaneous U$(1)_X$ breaking induces lepton masses according to  
\bea
\mu_N \sim \frac{m_{\rm SUSY}}{M_{Pl}} m_L,
\eea  
where $m_{\rm SUSY}$ is the supersymmetry breaking scale.
For $m_L < \Lambda_\ast = m_{\rm SUSY}$, one obtains the effective theory
given by eq.~(\ref{effective-model}) at energy scales below $m_{\rm SUSY}$. 
A large hierarchy between $\mu_N$ and $m_L$ can be generated naturally
because the lepton singlet mass additionally requires supersymmetry breaking.
For the ALP to be stable enough and constitute all of the dark matter, 
the mixing between $h$ and $\phi$ should be tiny, putting an upper bound on the supersymmetry 
breaking scale roughly as
\bea
m_{\rm SUSY} \lesssim
\frac{{10\, {\rm TeV}}}{yy^\prime \sin\alpha} 
\left(\frac{m_\phi}{10\,{\rm MeV}}\right)^{-3/2} \left(\frac{\tau}{10^{28}\,{\rm s}}\right)^{-1/2}, 
\eea
for $m_\phi< 2m_\mu$, where we have used the relations (\ref{longevity-1})
and (\ref{hidden-QCD-model}) for the confinement scale lying in the range (\ref{Confinement-condition}). 
Having a heavier mass, the ALP can decay to other SM particles, 
and the constraint becomes severer.
For instance,  
for the case that the ALP dominantly decays to $W$ bosons,   
the longevity condition (\ref{longevity-2}) requires
\bea
m_{\rm SUSY} \lesssim
\frac{10^{-13} \,{\rm TeV}}{yy^\prime\sin\alpha} 
\left(\frac{m_\phi}{10^3\, {\rm GeV}}\right)^{-1/2} \left(\frac{\tau}{10^{29}\,{\rm s}}\right)^{-1/2}, 
\eea
showing that $m_{\rm SUSY}$ above TeV requires small Yukawa couplings and/or
small CP phase leading to $yy^\prime \sin\alpha \lesssim 10^{-13}$
for $m_\phi$ around TeV.

Since SM gauge charged particles participate in generating the effective portal coupling, it is important 
to examine if the ALP is thermalized due to the interactions with them. 
If the reheating temperature of inflation $T_{\rm reh}$ is higher than $m_L$, the hidden lepton doublets 
are in thermal bath.
Because they are charged also under hidden QCD, the ALP would be thermalized via its anomalous coupling
to hidden QCD unless
\bea
T_{\rm reh} < T_{\rm dec}, 
\eea
where the ALP decoupling temperature is roughly given by the larger of the confinement scale 
$\Lambda_{\rm hid}$ and $10^4{\rm GeV}(f/10^9{\rm GeV})^2$.  
On the other hand, if $T_{\rm reh}$ is lower than $m_L$, hidden sector plasma can be colder 
than the SM plasma, and the ALP is never thermalized as long as
$T_{\rm dec}$ is higher than the hidden sector temperature after inflation.

Another issue to be considered is the cosmological effect of the $NN^c$ meson $\eta$,
which is integrated out to give the effective potential~(\ref{NP-Veff}).
The heavy meson has a mass of the order of $\Lambda_{\rm hid}$, and the relevant
interaction is   
\bea
\kappa \eta \phi h,
\eea
where the coupling constant is given by
\bea
\kappa \sim \frac{M^2 v}{f \Lambda_{\rm hid}}.
\eea 
where we have used that the mixing angle between the meson and Higgs is
roughly given by $\theta_{\rm mix} (m^2_h - m^2_\phi)f/\Lambda^3_{\rm hid}$,
and it is tiny in the region of parameter space of our interest. 
The model possesses an approximate $Z_2$ symmetry,  $\eta\to -\eta$ and $\phi\to -\phi$,
which is broken only slightly by nonzero $\alpha$ and thus can ensure that the ALP lives sufficiently long.
One may however wonder if the meson is a long-lived particle causing cosmological problems.
It would destroy light elements synthesized by BBN if decays during or after BBN,
or may alter the freeze-in production of ALP dark matter via the $\kappa$ interaction. 
These are simply avoided if the reheating temperature is lower than $\Lambda_{\rm hid}$
so that the meson is never in thermal equilibrium.

\subsection{Radiative Higgs Portal} 

Another interesting way to generate a Higgs portal interaction is to consider a slight breaking
of ALP shift symmetry that makes the ALP radiatively couple to the Higgs squared operator~\cite{Gupta:2015uea}. 
A simple model is obtained by adding vector-like lepton doublets $L+L^c$ and a lepton singlet $N$
with interactions respecting the ALP shift symmetry
\bea
\Delta {\cal L}
= m_L LL^c 
+ ye^{i\frac{\phi}{f} } HLN + y^\prime H^\dagger L^c N 
+ \frac{1}{2} \mu_{\rm s} e^{i(\frac{\phi}{f} + \alpha) } NN + {\rm h.c.},
\eea
where we have taken a field basis such that all the parameters are real and positive, which 
is always possible without loss of generality. 
Note that the first three interactions are responsible for an effective portal coupling, 
and in their presence the fourth term cannot be forbidden by the ALP shift symmetry.
The interactions depending on $\phi$ can arise, for instance, in a low energy effective 
theory below $f$ if ALP shift symmetry is linearly realized and spontaneously broken 
at a scale $f$.

Now we introduce a small mass term for the lepton singlet to break slightly the ALP shift
symmetry
\bea
\Delta {\cal L}_{\rm sb} = \frac{1}{2} \mu_{\rm sb} NN + {\rm h.c.},
\eea 
where $\mu_{\rm sb}$ is real and positive. 
Then, an effective potential is radiatively generated 
\bea
\Delta V_{\rm eff} = -M^2 \cos\left(\frac{\phi}{f} \right) |H|^2
- \frac{1}{16\pi^2}M^2\Lambda^2 \cos\left(\frac{\phi}{f}\right) 
- \mu^4_\phi \cos\left(\frac{\phi}{f} + \alpha \right),
\eea
with couplings given by
\bea
\label{radiative-portal}
M^2 &=& \frac{yy^\prime}{16\pi^2} \mu_{\rm sb} m_L 
\ln\left(\frac{\Lambda^2}{m^2_L}\right),   
\nonumber \\
\mu^4_\phi &=& \frac{1}{16\pi^2}\mu_{\rm sb} \mu_{\rm s} \Lambda^2.
\eea 
The potential term involving the CP phase $\alpha$ is sensitive to the cutoff scale $\Lambda$
as radiatively generated, differently from that in the case of a non-perturbatively generated portal.
For $\alpha\neq 0$, the ALP behaves like dark matter only when $\mu_{\rm s}$
is sufficiently small to suppress ALP-Higgs mixing.

The singlet mass term $\mu_{\rm sb}$ explicitly breaks the ALP shift symmetry 
at the perturbative level in the effective theory.
As a result, there arises an ALP coupling to the lepton doublets
\bea
\frac{yy^\prime}{16\pi^2}\mu_{\rm sb} \ln\left(\frac{\Lambda^2}{m^2_h}\right)
e^{i\frac{\phi}{f}} LL^c,
\eea 
at the loop level,
and thus the ALP interacts with electroweak gauge bosons via lepton doublet loops.
Such couplings make the ALP decay to gauge bosons, but can be suppressed 
if the shift symmetric mass of lepton doublets is sufficiently large.
In the parameter spaces (a) and (b) for freeze-in, the longevity condition requires
\bea
\mbox{(a)} &&
m_L \gtrsim 3\times 10^8\,{\rm GeV}
\left(\frac{M}{10^3\,{\rm GeV}}\right)^{1/2}
\left(\frac{m_\phi}{10\,{\rm MeV}}\right)^{5/8} \left(\frac{\tau}{10^{28}\,{\rm s}}\right)^{1/4},
\nonumber \\
\mbox{(b)} &&
m_L \gtrsim 4\times 10^{12}\,{\rm GeV}
\left(\frac{M}{10^3\,{\rm GeV}}\right)^{1/2}
\left(\frac{m_\phi}{10^3\,{\rm GeV}}\right)^{3/4} \left(\frac{\tau}{10^{29}\,{\rm s}}\right)^{1/4},
\eea
respectively, where we have used the relations (\ref{ALPmass1}) and (\ref{ALPmass2}).

To render the ALP stable further against $\alpha \neq 0$, we embed the interactions for a portal coupling 
in a supersymmetric model where the leptons acquire masses from the vacuum 
expectation value of $X$ after spontaneous breaking of U$(1)_X$. 
Let us assume that ALP shift symmetry is linearly realized by a superfield $\Phi$,
which implies $\Phi =\frac{f}{\sqrt2} e^{i\phi/f} +\cdots$ after spontaneous
U$(1)_\Phi$ breaking at $f$.  
We assign nonzero U$(1)_X$ charges only to $X$ and the leptons, and nonzero U$(1)_\Phi$
charges only to $\Phi$ and the leptons while making the lepton bilinear $LL^c$ neutral
so that the ALP does not have an anomalous coupling to SM gauge bosons.  
For instance, one can take the charge assignment
\bea 
\begin{tabular}{| c | c | c | c | c | c | c | c |} 
\hline
            & $H_u$ & $H_d$   & $L$   &  $L^c$    & $N$  & $X$   & $\Phi$         \\ \hline       
SU$(2)_L$  & ${\bf 2}$ & ${\bf \bar 2}$   & ${\bf 2}$      &  ${\bf \bar 2}$    & ${\bf 1}$  & ${\bf 1}$     & ${\bf 1}$ \\ \hline  
U$(1)_Y$   & $+1/2$ & $-1/2$  & $+1/2$              & $-1/2$             & $0$     & $0$          & $0$ \\ \hline  
U$(1)_X$  &0 & 0  & $-1/2$              & $-1/2$             & $+1/2$     & $+1$          & $0$ \\ \hline  
U$(1)_\Phi$  &0 & 0  & $-1/2$              & $+1/2$             & $-1/2$     & $0$          & $+1$ \\ \hline  
\end{tabular}
\nonumber 
\eea 
Then it is possible to obtain a supersymmetric embedding with
\bea 
K &\ni& \frac{X^*}{M_{Pl}}\Phi NN + {\rm h.c.},
\nonumber \\
W &\ni& XLL^c + \Phi H_d L N + H_u L^c N,
\eea
where we have omitted dimensionless coupling constants, and 
the cutoff scale of $\Phi$-dependent non-renormalizable operators, which is higher than $f$.
The lepton doublets become massive after U$(1)_X$ breaking, while the lepton singlet mass
parameter $\mu_s$ arises after supersymmetry and U$(1)_\Phi$ are broken further.
It thus follows  
\bea
\mu_{\rm s} = y \frac{m_{\rm SUSY}}{M_{Pl}} m_L,
\eea
with $m_L < \Lambda = m_{\rm SUSY}$,
which indicates that $\mu_s$ can be highly suppressed.
On the other hand, as the origin of the shift-symmetry breaking coupling $\mu_{\rm sb}$, one can consider 
a holomorphic operator $NN$ in the superpotential or $X^*NN$ in K\"ahler potential generated 
by non-perturbative effects breaking U$(1)_\Phi$ such as stringy instantons or field theoretic 
gaugino condensation.

Combining the relations (\ref{longevity-1}) and (\ref{longevity-2}) with 
the couplings (\ref{radiative-portal}),
one finds that the longevity condition is not sensitive to the supersymmetry breaking scale 
and is translated into 
\bea \label{rad_M1}
M \lesssim
4\,{\rm GeV}\,
\frac{y^\prime \ln(m_{\rm SUSY}/m_L)}{\sin\alpha}
\left(\frac{m_\phi}{10{\rm MeV}}\right)^{-3/2} \left(\frac{\tau}{10^{28}\,{\rm s}}\right)^{-1/2},
\eea
for $m_\phi <2m_\mu$ with the ALP mass roughly given by
$m_\phi \sim 1\,{\rm MeV}\times (m_{\rm SUSY}/{\rm TeV})^{4/5}$, while it leads to 
\bea \label{rad_M2}
M \lesssim
2\times 10^{-12}\,{\rm GeV}\,
\frac{y^\prime \ln(m_{\rm SUSY}/m_L)}{\sin\alpha}
\left(\frac{m_\phi}{10^3{\rm GeV}}\right)^{-1/2} \left(\frac{\tau}{10^{29}\,{\rm s}}\right)^{-1/2},
\eea
with $m_\phi \sim 380\,{\rm GeV}\times (m_{\rm SUSY}/10^9{\rm GeV})$
if the ALP decays mainly to $W$ bosons.  
In the above, the relation between the ALP mass and supersymmetry breaking scale 
assumes that the ALP makes up all of the dark matter of the universe. 
As discussed below eq.~(\ref{par_space}), the consistency condition requires
$M$ above TeV if the ALP is heavier than $m_h/2$.
The longevity bound (\ref{rad_M2}) would thus indicate that it is hard to realize a scenario of
radiative Higgs portal if the ALP is heavy.

Finally we discuss the condition for the ALP to remain unthermalized.
The interactions relevant to thermalization of the ALP read
\bea
\frac{\mu_s}{f} \phi NN +
y \frac{1}{f} \phi H L N + y^\prime H^\dagger L^c N,
\eea
where the second operator induces an effective 
Yukawa interaction but with a tiny coupling given by $yv/f$ 
after electroweak symmetry breaking.
Thus, if the reheating temperature is lower than the lepton doublet mass, 
the ALP never enters thermal equilibrium with the SM plasma. 
In the opposite case with $T_{\rm reh} > m_L$, the Yukawa coupling $\mu_s/f$ should
be smaller than about $10^{-7}$ to avoid thermalization of the ALP.

\section{Conclusions}
\label{conclusions}

An ALP coupled to the SM sector via a Higgs portal has recently been noticed to provide an explanation
for the puzzles in the SM such as the smallness of the electroweak scale and the origin
of baryon asymmetry. 
In this paper we have explored if it can solve the dark matter problem.  
The ALP is stabilized at a CP conserving vacuum and becomes stable if it has no other
non-derivative interactions. 
The perturbative shift symmetry then makes the ALP a natural and appealing candidate 
for freeze-in dark matter.

To UV complete the portal coupling, one can rely upon non-perturbative effects from
a hidden confining gauge group or radiative corrections from new leptons charged under
the shift symmetry.
UV models generally involve a CP violating interaction, which makes the ALP decay
into SM particles through mixing with the Higgs boson.
We found that such mixing is sufficiently suppressed in a natural way if embedded into
a supersymmetric theory. 
To avoid overclosure of the universe, ALP dark matter should be never in equilibrium with 
SM particles, constraining the properties of particles involved in generating an effective 
portal coupling.

\acknowledgments

We thank Kyu Jung Bae for helpful discussions. 
This work was supported by
the National Research Foundation of Korea (NRF) grant funded by the Korea
government (MSIP) (NRF-2018R1C1B6006061). 
SHI also acknowledges support from Basic Science Research Program through 
the National Research Foundation of Korea (NRF)
funded by the Ministry of Education (2019R1I1A1A01060680).


\bibliography{FZaxion_refs}
\bibliographystyle{utphys}

\end{document}